\documentclass[twocolumn,showpacs]{revtex4-2}

\usepackage{amsmath}
\usepackage{amssymb}
\usepackage{bm}
\usepackage{epsfig}
\usepackage{graphicx}
\usepackage{color}
\usepackage[colorlinks=true,citecolor=blue,urlcolor=black]{hyperref}
\usepackage[english]{babel}

\renewcommand{\Re}{\mathop{\rm Re}}
\renewcommand{\Im}{\mathop{\rm Im}}

\renewcommand{\S}{\mathop{\mathcal S}}
\newcommand{\X}{\mathop{\mathcal X}}
\newcommand{\Y}{\mathop{\mathcal Y}}
\newcommand{\Z}{\mathop{\mathcal Z}}

\begin{document}

\title{Exciton fine structure splitting and linearly polarized emission in strained transition-metal dichalcogenide monolayers}

\author{M. M. Glazov}

\affiliation{Ioffe Institute, 194021 Saint Petersburg, Russia}
\affiliation{National Research University, Higher School of Economics, 190121 Saint Petersburg, Russia}

\author{Florian Dirnberger}
\affiliation{Department of Physics, City College of New York, New York, New York 10031, USA}
\affiliation{Department of Physics, The Graduate Center, City University of New York, New York, New York 10016, USA}
\affiliation{Integrated Center for Applied Physics and Photonic Materials, Institute of Applied Physics and W\"urzburg-Dresden Cluster of Excellence ct.qmat, Technische Universit\"at Dresden, 01062 Dresden, Germany}

\author{Vinod M. Menon}
\affiliation{Department of Physics, City College of New York, New York, New York 10031, USA}
\affiliation{Department of Physics, The Graduate Center, City University of New York, New York, New York 10016, USA}

\author{Takashi Taniguchi}
\affiliation{International Center for Materials Nanoarchitectonics, National Institute for Materials Science, Tsukuba, Ibaraki 305-004, Japan}
\author{Kenji Watanabe}

\affiliation{Research Center for Functional Materials, National Institute for Materials Science, Tsukuba, Ibaraki 305-004, Japan}

\author{Dominique Bougeard}
\affiliation{Department of Physics, University of Regensburg, 93040 Regensburg, Germany}

\author{Jonas D. Ziegler}
\author{Alexey Chernikov}
\affiliation{Integrated Center for Applied Physics and Photonic Materials, Institute of Applied Physics and W\"urzburg-Dresden Cluster of Excellence ct.qmat, Technische Universit\"at Dresden, 01062 Dresden, Germany}

\date{\today}

\begin{abstract}
We study theoretically effects of an anisotropic elastic strain on the exciton energy spectrum fine structure and optical selection rules in atom-thin crystals based on transition-metal dichalcogenides. The presence of strain breaks the chiral selection rules at the $\bm K$-points of the Brillouin zone and makes optical transitions linearly polarized. The orientation of the induced linear polarization is related to the main axes of the strain tensor. Elastic strain provides an additive contribution to the exciton fine structure splitting in agreement with experimental evidence obtained from uniaxially strained WSe$_2$ monolayer.
The applied strain also induces momentum-dependent Zeeman splitting. Depending on the strain orientation and magnitude, Dirac points with a linear dispersion can be formed in the exciton energy spectrum. We provide a symmetry analysis of the strain effects and develop a microscopic theory for all relevant strain-induced contributions to the exciton fine structure Hamiltonian. 
\end{abstract}

\maketitle

\section{Introduction}

Elastic deformations and strain strongly affect electronic, excitonic and optical properties of solids~\cite{birpikus_eng}. Deformation effects turn out to be especially important for two-dimensional (2D) materials where strain can be applied locally~\cite{Castellanos-Gomez:2013tn} providing significant tunability of the band parameters and optical response of semiconducting transition-metal dichalcogenide monolayers (TMDC MLs)~\cite{Rostami15,PhysRevB.100.195126}. Such systems with MoS$_2$, MoSe$_2$, WS$_2$, and WSe$_2$ being their most prominent representatives are actively studied nowadays owing to their exceptional optical properties, intriguing fundamental physics, and wide prospects of applications~\cite{Xiao:2012cr,Kolobov2016book,RevModPhys.90.021001}. Deformations can be used to control exciton transition energy and linewidths~\cite{Zhu:2013ve,2053-1583-2-1-015006,2053-1583-3-2-021011,Phuc:2018aa,PhysRevB.98.115308,Khatibi_2018,Niehues:2018tm,Benimetskiy:2019uz}, scattering and diffusion~\cite{Rosati_2020}. Inhomogeneous strain allows one to confine~\cite{PhysRevB.97.195454,Smiri:2021wv} and steer excitons owing to the funneling effect~\cite{Moon:2020vw,Hyeongwoo:vi,Bai:2020uh,Rosati:2021wt,PhysRevB.104.085405,Florian:ub}.

In atom-thin TMDCs bright excitons occupy two valleys $\bm K_+$ and $\bm K_-$ and emit circularly polarized light. Valley-tagged excitons can be considered as a prototypical two-level system with the valley polarization and coherence being mapped to the exciton pseudospin-$1/2$~\cite{glazov2014exciton,PSSB:PSSB201552211}. Valley orientation of excitons by circularly polarized light and formation of valley coherence under linearly polarized excitation~\cite{Kioseoglou,Zeng:2012ys,Sallen:2012qf,Mak:2012qf,Jones:2013tg,PSSB:PSSB201552211,PhysRevLett.117.187401,app8071157,Lundt:2019aa,Glazov2020b,Glazov_2021} is among the most promising effects for applications of TMDC MLs in quantum technologies. Electron-hole exchange interaction mixes excitons in the $\bm K_\pm$ valleys and acts as an effective magnetic field -- synthetic spin-orbit coupling -- making it possible to manipulate valley polarization~\cite{Yu:2014fk-1,PhysRevB.89.205303,glazov2014exciton,PSSB:PSSB201552211} akin to the manipulation of electron spins via spin-orbit coupling.~\cite{dresselhaus55,rashbasheka,rashba64,Rashba03,golub_ganichev_BIASIA,dyakonov_book}. Electron-hole exchange interaction governs the fine structure of excitonic states~\cite{Wang:2017b,Durnev_2018,RevModPhys.90.021001} and also causes valley depolarization of excitons ~\cite{glazov2014exciton,Yu:2014fk-1,PhysRevB.89.205303,PhysRevB.90.161302,PSSB:PSSB201552211,prazdnichnykh2020control} similarly to the spin depolarization of excitons in quantum wells~\cite{maialle93,goupalov98,ivchenko05a}. 

Here we demonstrate theoretically and confirm experimentally that the strain strongly affects the bright exciton fine structure in TMDC MLs. It breaks the three-fold rotational symmetry of a monolayer and gives rise to a splitting of a radiative doublet into the states linearly polarized along the main axes of the strain tensor. The strain results in overall linear polarization of the emission of 2D TMDC. We also predict a combined effect of the strain and exciton motion which leads to the linear-in-wavevector splitting of the circularly polarized exciton states. We present symmetry analysis of the strain effects in Sec.~\ref{sec:sym} and microscopic mechanisms of the effects in Sec.~\ref{sec:micro}. An experimental demonstration of the interdependence of exciton energy shifts and fine structure splitting under strain is provided in Sec.~\ref{sec:exper}. Section~\ref{sec:concl} contains the conclusion and outlook.

\section{Symmetry analysis}\label{sec:sym}

In this section we employ symmetry arguments to derive an effective Hamiltonian of the exciton radiative doublet in TMDC monolayers in the presence of elastic strain, Sec.~\ref{subsec:Heff}. Further, in Sec.~\ref{subsec:energ} we analyze the energy spectrum and discuss the formation of the Dirac points with conical dispersion.

\subsection{Effective Hamiltonian}\label{subsec:Heff}

\begin{figure*}[t]
\includegraphics[width=0.7\linewidth]{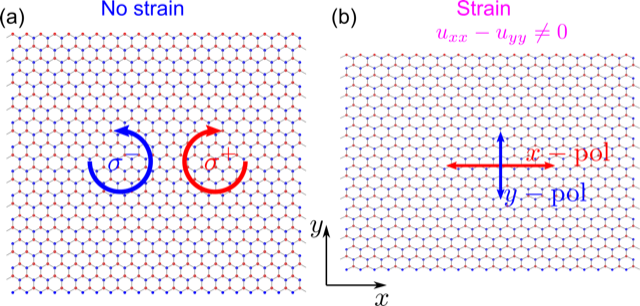}
\caption{Schematic top view of a TMDC ML in the absence (a) and in the presence (b) of strain, $u_{xx} \ne u_{yy}$, $u_{xy}=0$. }\label{fig:1}
\end{figure*}

We consider the fine structure of the radiative doublet of exciton states in the presence of strain in the 2D TMDC. Let $\bm u(\bm r)$ be the local displacement vector of the atoms in the monolayer. For small deformations considered hereafter the strain tensor reads ($i,j=x,y,z$ denote the Cartesian components)
\begin{equation}
\label{strain}
u_{ij} \equiv \frac{1}{2} \left( \frac{\partial u_i}{\partial x_j} + \frac{\partial u_j}{\partial x_i}\right) = u_{ji}.
\end{equation}
To obtain an effective Hamiltonian of the exciton in the presence of strain, we use the method of invariants~\cite{birpikus_eng,ivchenkopikus}. We consider a class of TMDC monolayers with $D_{3h}$ point symmetry that includes MoS$_2$, MoSe$_2$, MoTe$_2$, WS$_2$ and WSe$_2$ as the most prominent examples, our analysis applies to all these systems.
In the $D_{3h}$ point group of the monolayer the following strain tensor components correspond to the irreducible representations:
\begin{enumerate}
\item $A_1'$ (invariant): $u_{zz}$; $u_{xx}+u_{yy}$.
\item $E'$: ($2u_{xy}$, $u_{xx} - u_{yy}$,).
\item $E''$: ($u_{xz}$, $u_{yz}$).
\end{enumerate}
The bright excitonic states -- the radiative doublet -- transform as the in-plane coordinates, i.e., according to the $E'$ irreducible representation.  Taking into account that $E'\otimes E' = A_1'+A_2'+E'$, we obtain the following expression for the effective Hamiltonian of the radiative doublet in the presence of strain:
\begin{equation}
\label{H:eff}
\mathcal H = \Xi (u_{xx} + u_{yy}) \hat I + \Xi' u_{zz} \hat I+ \frac{\hbar^2 K^2}{2M} \hat I + \frac{\hbar}{2} (\hat{\bm \sigma} \cdot \bm \Omega).
\end{equation}
Here $\hat I$ is the $2\times 2$ unit matrix, and the contributions with $\propto u_{xx} + u_{yy}$ and $u_{zz}$ describe the overal shift of the exciton energy, $\Xi$ and $\Xi'$ are the corresponding deformation potentials. These contributions are responsible for the exciton energy tuning by elastic strain~\cite{2053-1583-2-1-015006,2053-1583-3-2-021011,Phuc:2018aa}, electron- and exciton-acoustic phonon scattering~\cite{PhysRevB.90.045422,PhysRevB.85.115317,Song:2013uq,PhysRevB.94.205423,Selig:2016aa,PhysRevLett.119.187402,shree2018exciton}, and exciton funneling~\cite{Moon:2020vw,Hyeongwoo:vi,Florian:ub}. The kinetic energy $\hbar^2 K^2/2M$ is presented disregading the strain with $\bm K = (K_x,K_y)$ being the two-dimensional exciton wavevector; generally, in the presence of strain the kinetic energy acquires the terms $\propto K_iK_j u_{ij}$ which describe the strain-induced renormalization of the effective mass and its anisotropy. The last term in Eq.~\eqref{H:eff} describes the exciton fine structure. Here $\hat {\bm \sigma} = (\hat \sigma_x,\hat \sigma_y, \hat \sigma_z)$ is the vector composed of the Pauli matrices describing the exciton pseudospin, with $\hat \sigma_z$ describing the splitting of exciton states in circular polarizations and $\hat\sigma_x$, $\hat\sigma_y$ the splitting into the linearly polarized components~\cite{ivchenko05a, glazov2014exciton}, and $\bm \Omega$ is the effective magnetic field acting on the exciton pseudospin:
\begin{subequations}
\label{Omega}
\begin{align}
&\Omega_x = \mathcal A(K) (K_x^2-K_y^2) + \mathcal B (u_{xx} - u_{yy}),\\
&\Omega_y = 2\mathcal A(K)K_x K_y + 2\mathcal B u_{xy},\\
&\Omega_z=\mathcal C[(u_{xx} - u_{yy}) K_x - 2u_{xy} K_y].
\end{align}
\end{subequations}
Note that the exciton pseudospin Pauli matrices $\hat\sigma_x$ and $\hat\sigma_y$ are even at the time reversal, while the matrix $\hat\sigma_z$ is odd at the time-reversal.

The product  $\mathcal A(K)K^2$ describes the longitudinal-transversal splitting of the excitonic states~\cite{glazov2014exciton,Yu:2014fk-1,PhysRevB.89.205303,PhysRevB.101.115307,prazdnichnykh2020control}. The parameter $\mathcal B$ describes the effect of strain, and the parameter $\mathcal C$ describes an effective magnetic field arising due to the exciton propagation in the presence of strain. The parameter $\mathcal C \ne 0$ appears due to the lack of the inversion symmetry in the system. Here we use the same set of axes as in Ref.~\cite{PhysRevB.95.035311} where one of the three vertical reflection planes on $D_{3h}$ point group $\sigma_v \parallel (yz)$, see Fig.~\ref{fig:1}, and the components $(K_x, K_y)$ transform as $(2u_{xy},u_{xx} - u_{yy})$.

Note that the mixed components of the strain, $u_{xz}=u_{zx}$ and $u_{yz}=u_{zy}$, mix spin-forbidden dark (also called gray) $z$-polarized exciton~\cite{Wang:2017b} with the in-plane polarized optically active states. Such effects are beyond the scope of the present work. 

\subsection{Analysis of the energy spectrum}\label{subsec:energ}

The strain breaks the three-fold rotation symmetry of the 2D TMDC, see Fig.~\ref{fig:1}. In the absence of strain the axes $x$ and $y$ are equivalent from the viewpoint of linear optics and the eigenpolarizations can be selected to be circular, $\sigma^+$ and $\sigma^-$, Fig.~\ref{fig:1}(a). In the presence of strain, the in-plane axes of the structure are no longer equivalent, Fig.~\ref{fig:1}(b), and the eigenpolarizations are linear ones along the main axes of the strain. The eigenstates of the exciton at rest ($K=0$) are linearly polarized along the main axes of the strain tensor $x'$ and $y'$, and the splitting is given by $\hbar \mathcal B  (u_{x'x'} - u_{y'y'})$. Note that in the main axes the off-diagonal components of the strain tensor are absent, $u_{x'y'} = u_{y'x'} \equiv 0$; in Fig.~\ref{fig:1}(b) the main axes are $x$ and $y$ (generally the main axes $x'$ and $y'$ do not have to match $x$ and $y$ axes of the lattice in Fig.~\ref{fig:1}). The splitting is linear in the strain for small deformations.

\begin{figure}[t]
\includegraphics[width=0.8\linewidth]{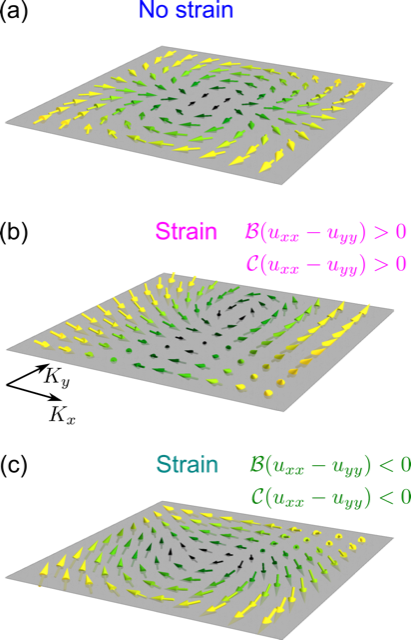}
\caption{Illustration of vector field $\bm \Omega$ (arrows) as function of the exciton wavevector $\bm K$.  (a) Strain is absent, the effective magnetic field results solely from the exciton longitudinal-transverse splitting. (b,c) Strain is present: $u_{xy}=0$, $u_{xx} - u_{yy}$ has different signs for the panels (b) and (c). }\label{fig:2}
\end{figure}

The pseudomagnetic field $\bm \Omega$ as function of the wavevector $\bm K$ is schematically shown in Fig.~\ref{fig:2}. Three panels show the case without strain (a) and with strain of opposite sign (b,c): Here we took 
\begin{equation}
\label{main:xy}
u_{xy}=0, \quad\mbox{and} \quad u_{xx} - u_{yy} \ne 0.
\end{equation} 
In the case (a), no strain is present, the field $\bm \Omega$ lies in the $(xy)$ plane and its components are described by the second angular harmonics~\cite{glazov2014exciton,PSSB:PSSB201552211}. The magnitude of $\bm \Omega$ depends only on the absolute value of $\bm K$. Strain breaks the axial symmetry of the system making the magnitude $\Omega$ dependent on both the magnitude and direction of $\bm K$, as shown in panels (b) and (c). Also, a $z$-component of $\bm \Omega$ appears. The reversal of the sign of the strain changes the distribution of the field $\bm \Omega$, cf. (b) and (c).

\begin{figure}[t]
\includegraphics[width=0.49\linewidth]{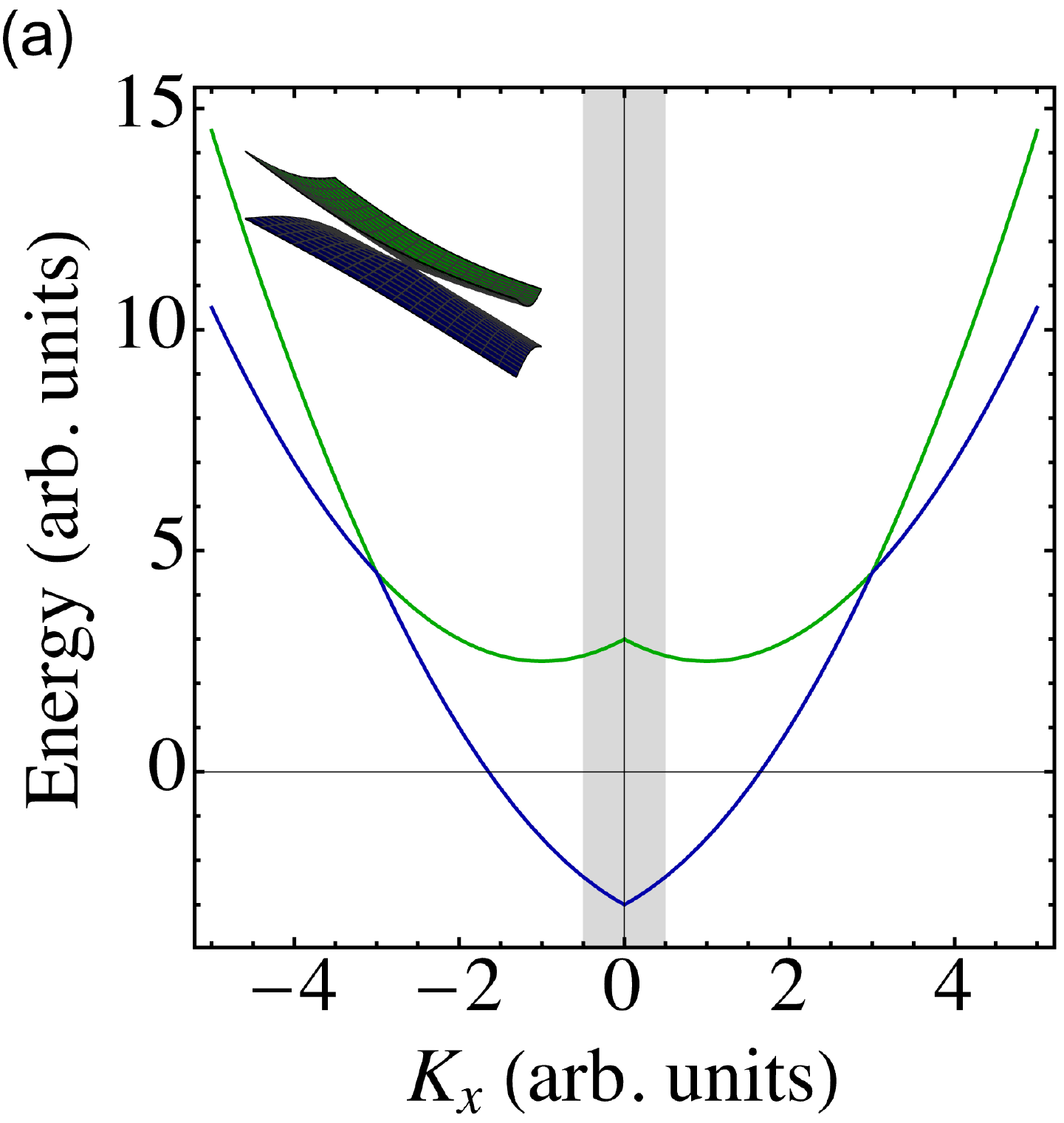}
\includegraphics[width=0.49\linewidth]{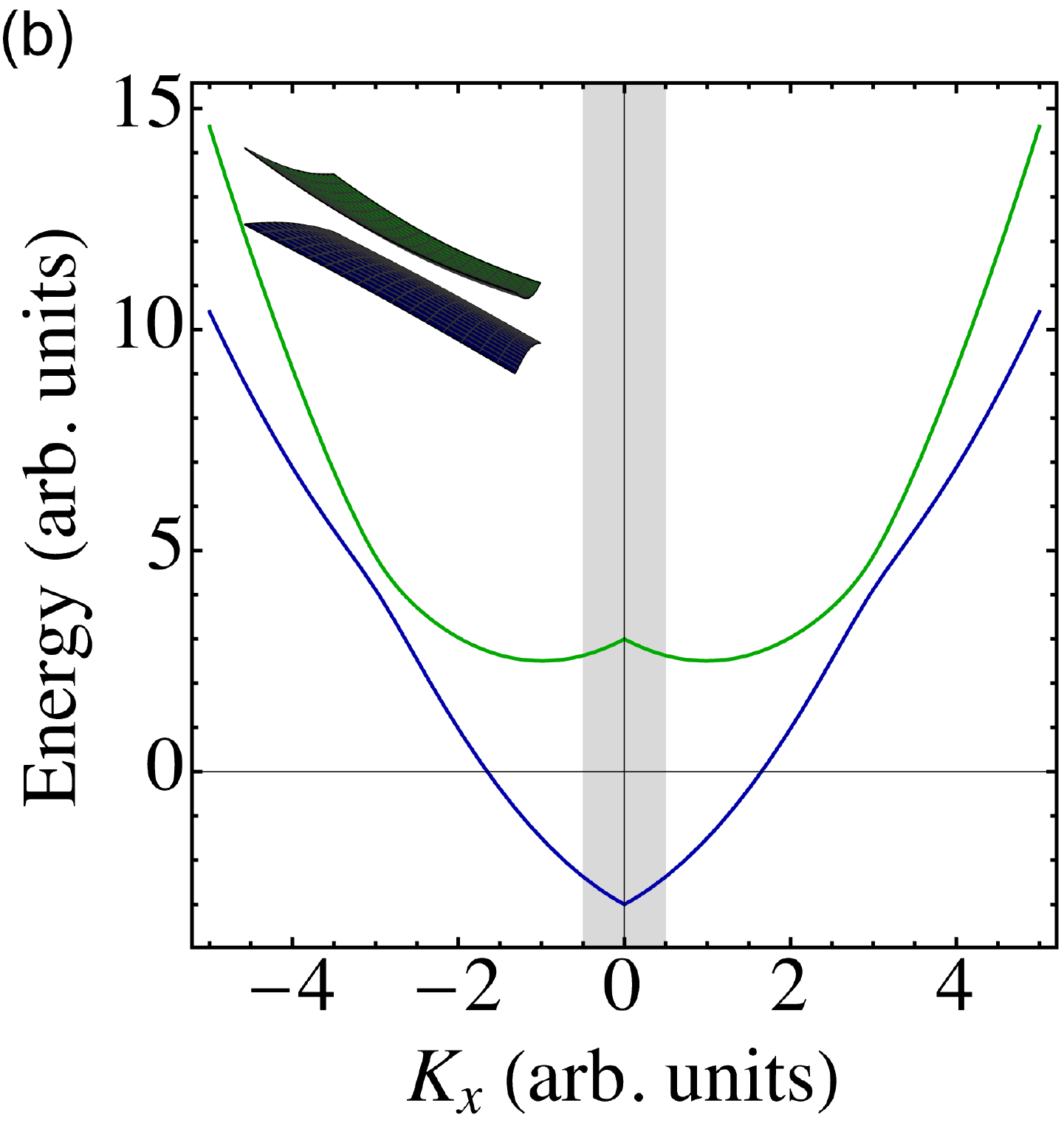}
\caption{Illustration of the exciton dispersion found by diagonalization of the Hamiltonian~\eqref{H:eff} as function of $K_x$ at $K_y=0$ for $K_*^2<0$ ($\mathcal B(u_{xx} - u_{yy})/\mathcal A<0$) at $\mathcal C=0$ (a) and $\mathcal C \ne 0$ (b). Insets show three-dimensional plots of the dispersion in the vicinity of the Dirac point $K_x =- \sqrt{|K_*^2|}$, $K_y=0$, Eq.~\eqref{Ky=0}. The light cone is shown by the shaded area.}\label{fig:3}
\end{figure}

Depending on the strain the energy spectrum of excitons can possess Dirac points where the polarization eigenstates are degenerate and the energy spectrum is conical. For illustrative properties let us assume that the main axes of the strain correspond to the $(xy)$ axes of the structure, namely, Eq.~\eqref{main:xy} is fulfilled. In the centrosymmetric approximation where $\mathcal C=0$ the effective magnetic field $\bm \Omega$ vanishes at 
\begin{subequations}
\label{Dirac:susp}
\begin{equation}
\label{Kx=0}
K_x=0, \quad K_y = \pm \sqrt{K_*^2}, \quad K_*^2>0,
\end{equation}
or at
\begin{equation}
\label{Ky=0}
K_x= \pm \sqrt{-K_*^2},\quad K_y=0 \quad K_*^2<0,
\end{equation}
\end{subequations}
where the parameter $K_*$ is determined from the equation.
\begin{equation}
\label{K*}
K_*^2 = \frac{\mathcal B(u_{xx} - u_{yy})}{\mathcal A(|K_*|)}.
\end{equation}
The points~\eqref{Kx=0} and \eqref{Ky=0} in the $\bm K$-space correspond to a vanishing exciton polarization splitting, and the effective Hamiltonian in the vicinity of each point takes the form
\begin{subequations}
\label{H:Dirac:susp}
\begin{equation}
\label{H:Kx=0}
\mathcal H(\tilde{\bm K}) \approx \frac{\hbar^2 |K_*|^2}{2M} \pm \hbar \mathcal V_*(\sigma_y \tilde K_x - \sigma_x\tilde K_y),
\end{equation}
or
\begin{equation}
\label{H:Ky=0}
\mathcal H(\tilde{\bm K}) \approx \frac{\hbar^2 |K_*|^2}{2M} \pm \hbar \mathcal V_*(\sigma_x \tilde K_x + \sigma_y\tilde K_y),
\end{equation}
\end{subequations}
depending on whether Eq.~\eqref{Kx=0} or \eqref{Ky=0} is fulfilled.
Here $\tilde{\bm K}$ is the wavevector reckoned from the corresponding Dirac point, Eq.~\eqref{Dirac:susp}, and the effective velocity is given by 
\begin{equation}
\label{V*}
\mathcal V_* = \hbar |K_*| \mathcal A(|K_*|).
\end{equation}
In Eqs.~\eqref{H:Dirac:susp} we have omitted $\sigma_{x,y}$ independent $\tilde{\bm K}$-linear terms $\pm \hbar^2 \tilde K_{x,y} |K_*|/M$ stemming from the parabolic exciton dispersion, which result in a tilt of the Dirac cones. The dispersion calculated using the total Hamiltonian~\eqref{H:eff} at $\mathcal C=0$ for $K_*^2<0$ is shown in Fig.~\ref{fig:3}(a). It clearly shows the Dirac points with tilted conical dispersion.

 For vanishing strain $|K_*|\to 0$ and Dirac points enter the light cone ($|K_*|<\omega_0/c$, where $\omega_0$ is the exciton emission frequency shown by the gray shading in Fig.~\ref{fig:3}) where excitons radiatively decay and $\mathcal A(K)$ becomes imaginary, see Eq.~\eqref{alpha} and Ref.~\cite{glazov2014exciton} for details. For the same reason, the dispersion in Fig.~\ref{fig:3} demonstrates non-analytical features at $K_x=0$. In such a case, there is no sense to discuss Dirac cones due to significant radiative damping of the excitons.

Interestingly, the linear dispersion in Eqs.~\eqref{H:Dirac:susp} can be strongly modified if the lack of the inversion symmetry is taken into account, i.e., where $\mathcal C \ne 0$. If the parameters of the strain are such that $K_*^2>0$ and Eq.~\eqref{Kx=0} is fulfilled, then $\Omega_z$ vanishes at $K_x=0$, $K_y = \pm K_*$, and the dispersion acquires an additional $\tilde K_x$-linear contribution $\propto \sigma_z \mathcal C(u_{xx} - u_{yy}) \tilde K_x$ that simply results in a `tilt' of the eigenstates and anisotropy of the Dirac cone. By contrast, if $K_*^2<0$ such that Eq.~\eqref{Ky=0} is fulfilled, then a gap opens in the spectrum: Indeed, the additional $\mathcal C$-linear contribution to the effective Hamiltonian~\eqref{H:Ky=0} is given by
\begin{equation}
\delta \mathcal H = \pm \sigma_z \frac{\hbar}{2} \mathcal C (u_{xx} - u_{yy})\sqrt{-K_*^2},
\end{equation} 
and it does not depend on the wavevector to the lowest order. Thus, the gap is given by
\[
E_G = \frac{\hbar}{2}|\mathcal C (u_{xx} - u_{yy})|\sqrt{-K_*^2}.
\]
Such situation is illustrated in Fig.~\ref{fig:3}(b) where the gap is seen. Note that for arbitrary orientation of the strain with respect to the crystalline axes $x$ and $y$ the gap is generally non-zero, it closes  for specific orientations of the strain only. 


\section{Microscopic mechanisms}\label{sec:micro}

Here we present the microscopic model of the strain-induced contributions to the effective magnetic field $\bm\Omega$ acting on the exciton pseudospin. We start the analysis with the strain induced effects on the single-particle states and optical selection rules, Sec.~\ref{subsec:single}, then address the excitonic effects demonstrating the role of the long-range, Sec.~\ref{subsec:lr}, and short-range, Sec.~\ref{subsec:sr}, exchange interaction. In Sec.~\ref{subsec:ncs} we address the effects related to the lack of the inversion center. It is noteworthy that all common TMDC MLs have similar bandstructure. Accordingly, the mechanisms described below are rather general and not limited to any specific combination of transition metal and chalcogen.

\subsection{Strain-induced band mixing and optical selection rules}\label{subsec:single}

In order to describe the strain effects on the single-particle Bloch functions, we resort to the 4-band $\bm k\cdot \bm p$-model of the TMDC ML bandstructure and, in addition to the conduction $c$ and valence $v$ bands, also include the remote bands $c+2$ and $v-3$ into the consideration (see Refs.~\cite{2053-1583-2-2-022001,Durnev_2018,PhysRevB.95.035311} for details) whose Bloch functions at the $\bm K_+$ point of the Brillouin zone can be chosen as 
\begin{equation}
\label{bands}
|c+2\rangle^+ = \frac{\X'-\mathrm i\Y'}{\sqrt{2}}, \quad |c\rangle^+ = \frac{\X+\mathrm i\Y}{\sqrt{2}},
\end{equation}
\[
|v\rangle^+ = \S, \quad |v-3\rangle^+ = \frac{\X''-\mathrm i\Y''}{\sqrt{2}}.
\]
In the $\bm K_-$ valley the functions are obtained by the time-reversal transformation and can be chosen, e.g., as complex conjugates of Eq.~\eqref{bands}; Bloch functions $\X$, $\Y$, $\Z$ transform as the corresponding Cartesian coordinate, $\S$ is the invariant function; we use primes to denote different orbital composition of the Bloch functions in different bands. Since we are interested in the fine structure of the bright excitonic states, we consider the Bloch states with the fixed spin component, i.e, $\uparrow$ in the $\bm K_+$ valley and $\downarrow$ in the $\bm K_-$ one. We recall that the bands $c+1$, $v-1,v-2$ are odd at $z\to -z$ reflection and provide no contribution to the parameter $\mathcal B$. Such a model adequately describes the band structure and symmetry of the transition metal dichalcogenide monolayers in the vicinity of the $\bm K_\pm$ points~\cite{2053-1583-2-2-022001}. The  $\bm k\cdot \bm p$ Hamiltonian of the $\bm K_+$ valley reads~\cite{2053-1583-2-2-022001,Durnev_2018,PhysRevB.95.035311}
\begin{equation}
\label{H4+}
\mathcal H_+ = \begin{pmatrix}
E_{c+2} & \gamma_6 k_- & \gamma_4 k_+ & 0 \\
\gamma_6^* k_+ & E_c & \gamma_3 k_- & \gamma_5 k_+ \\
\gamma_4^* k_- & \gamma_3^* k_+ & E_v & \gamma_2 k_-\\
0 & \gamma_5^* k_- & \gamma_2^* k_+ & E_{v-3}
\end{pmatrix},
\end{equation}
where $k_\pm = k_x \pm \mathrm i k_y$, $E_n$ ($n=v-3,v,c,c+2$) are the energies of the band edges and $\gamma_{2,\ldots, 6}$ are the parameters proportional to the interband momentum matrix elements. The Hamiltonian in the $\bm K_-$ valley, $\mathcal H_-$, is obtained by the conjugation of $\mathcal H_+$.

The strain mixes the bands in Eq.~\eqref{bands}. In what follows we disregard the contributions $u_{zz}$, $u_{xz}$, $u_{yz}$ that involve the $z$-component of atomic displacements and assume that only $u_{ij}$ with $i,j=x,y$ are non-zero. Following the general approach of Ref.~\cite{birpikus_eng} we derive the following strain-induced contribution to the $\bm k\cdot \bm p$-Hamiltonian in the $\bm K_+$ valley:
\begin{equation}
\label{H4+:s}
\mathcal H_+^{(s)} = \begin{pmatrix}
{\Xi_{c+2}}u_0 & \Xi_{c+2,c} u_{+2} & \Xi_{c+2,c} u_{-2} & \Xi_{c+2,v-3}u_0 \\
\Xi_{c,c+2} u_{-2} & \Xi_c u_0 & \Xi_{c,v} u_{+2} & \Xi_{c,v-3} u_{-2}\\
\Xi_{c+2,v}u_{+2} & \Xi_{v,c} u_{-2} & \Xi_v u_0 & \Xi_{v,v-3} u_{+2}\\
\Xi_{v-3,c+2}u_0 & \Xi_{v-3,c} u_{+2} & \Xi_{v-3,v} u_{-2} & \Xi_{v-3}u_0
\end{pmatrix},
\end{equation}
where we introduced the notations
\begin{equation}
\label{not}
u_{\pm 2} = u_{xx} - u_{yy} \pm 2\mathrm i u_{xy}, \quad u_0 = u_{xx} + u_{yy},
\end{equation}
and the interband deformation potential matrix elements $\Xi_{n,n'} = \Xi_{n',n}^*$ ($n,n'=c,v,c+2,v-3$) and also used a short-hand notation $\Xi_{n}$ for $\Xi_{n,n}$. The parameters $\Xi_{n,n'}$ should be found from atomistic calculations similar to those in Ref.~\cite{PhysRevB.100.195126}. Typically, $\Xi_{n,n'}$ are in the range from units to tens of eV.

In the first order in $u_{ij}$ we obtain at $\bm k=0$
\begin{widetext}
\begin{subequations}
\label{Bloch:strain}
\begin{align}
|\tilde c\rangle^+ = |c\rangle^+ + u_{+2} \left(\frac{\Xi_{c+2,c}}{E_{c} - E_{c+2}} |c+2\rangle^+ + \frac{\Xi_{v-3,c}}{E_{c} - E_{v-3}} |v-3\rangle^+ \right) + u_{-2} \frac{\Xi_{v,c}}{E_{c} - E_{v}}|v\rangle^+,\\
|\tilde v\rangle^+ = |v\rangle^+ + u_{-2} \left(\frac{\Xi_{c+2,v}}{E_{v} - E_{c+2}} |c+2\rangle^+ + \frac{\Xi_{v-3,v}}{E_{v} - E_{v-3}} |v-3\rangle^+ \right) + u_{+2} \frac{\Xi_{c,v}}{E_{c} - E_{v}}|v\rangle^+.
\end{align}
\end{subequations}
\end{widetext}
 The expressions for the Bloch states in the $\bm K_-$ valley can be obtained from Eqs.~\eqref{Bloch:strain} by the complex conjugation.  The ratios $\Xi_{n,n'}/(E_{n} - E_{n'})$ can be estimated as $1\ldots 10$.

Equations~\eqref{Bloch:strain} demonstrate that the selection rules for the optical transitions at the $\bm K_\pm$ points change and the strict (chiral) selection rules no longer hold in the presence of the in-plane strain. Making use of the 4-band $\bm k\cdot \bm p$-Hamiltonian~\eqref{H4+}, we calculate the interband velocity matrix element for the transitions in $\sigma^\pm$ polarizations, $ \tilde{ \bm v}_{cv}^+ = \langle \tilde c |\hbar^{-1} \bm \nabla_{\bm k} \mathcal H_+|\tilde v\rangle$, to be
\begin{equation}
\label{vcv+}
\tilde v_{cv}^+(\sigma^+) = \frac{\gamma_3}{\hbar}, \quad \tilde v_{cv}^+(\sigma^-) =u_{-2} \eta \frac{\gamma_3}{\hbar},
\end{equation}
where the parameter $\eta$ equals to
\begin{multline}
\label{eta}
\eta= \frac{1}{\gamma_3}\left(\frac{\Xi_{c,c+2}}{E_c - E_{c+2}} \gamma_4 + \frac{\Xi_{c,v-3}}{E_c - E_{v-3}}\gamma_2^* + \right.\\
\left. \frac{\Xi_{c+2,v}}{E_{v} - E_{c+2}} \gamma_6^* + \frac{\Xi_{v-3,v}}{E_{v} - E_{v-3}}  \gamma_5\right).
\end{multline}
Similarly, transition matrix elements in the opposite ($\bm K_-$) valley take the form
\begin{equation}
\label{vcv-}
\tilde v_{cv}^-(\sigma^+) = u_{2} \eta{\frac{\gamma_3^*}{\hbar}}, \quad \tilde v_{cv}^-(\sigma^-) ={\frac{\gamma_3^*}{\hbar}}.
\end{equation}
One can check that for a consistent choice of the phases of the Bloch functions in Eq.~\eqref{bands} the parameter $\eta$ is real.

Equations~\eqref{vcv+} and \eqref{vcv-} demonstrate that the optical transitions are, in the presence of strain, elliptically polarized. The strain-induced linear polarization is along the main axes of the strain tensor $u_{ij}$ and the same in both valleys. In what follows we assume that $u_{xy} = u_{yx}=0$, i.e., the axes $(xy)$ are the main axes, Fig.~\ref{fig:1}(b), and obtain from Eqs.~\eqref{vcv+} and \eqref{vcv-}
\begin{equation}
\label{Pl}
P_l =\frac{|\tilde v^\pm_{cv,x}|^2 - |\tilde v^\pm_{cv,y}|^2}{|\tilde v^\pm_{cv,x}|^2 + |\tilde v^\pm_{cv,y}|^2} =  2 (u_{xx} - u_{yy}) \eta.
\end{equation}
Thus, interband emission of strained ML becomes linearly polarized with the overal (spectrally integrated) polarization degree given by Eq.~\eqref{Pl}. For rough estimates one can take $\eta \sim 1\ldots 10$ yielding $P_l$ in the range of several to tens of percent per $\%$ strain.

Note that the strain itself, at $\bm k=0$, cannot result in the splitting of the non-interacting electron and hole valley-degenerate states. This is because of the Kramers theorem resulting from the time-reversal symmetry: The single-particle valley splitting is provided only by the time-reversal non-invariant perturbations (e.g., magnetic field), while the strain does not change its sign at the time reversal as clearly seen from the definition~\eqref{strain}. Thus, in order to derive $\mathcal B\ne 0$, one has to take into account both the strain-induced modification of the band structure and the electron-hole interaction. Since the exchange interaction contains both the long- and short-range contributions the parameter $\mathcal B = \mathcal B_{lr} + \mathcal B_{sr}$, where the subscripts $lr$ and $sr$ denote, respectively, the long-range and short-range effects. We discuss these effects in the following subsections.

\subsection{Long-range exchange interaction}\label{subsec:lr}

The long-range contribution to the electron-hole exchange interaction can be interpreted as a process of (virtual) recombination and generation process of the electron-hole pair~\cite{BP_exch71,zhilich72:eng,denisovmakarov,goupalov98,goupalov:electrodyn,glazov2014exciton,prazdnichnykh2020control,2022arXiv220310295G}. It allows us to consider the long-range exchange within the purely electrodynamical approach taking into account the coupling of exciton with induced radiation, see Ref.~\cite{glazov2014exciton,prazdnichnykh2020control} for detailed analysis of the effect for TMDC MLs.

The modification of selection rules provides, via the long-range exchange interaction, the coupling between $\sigma^+$ and $\sigma^-$ excitons already at $K=0$. For a free-standing monolayer this coupling is dissipative (cf. Ref.~\cite{PhysRevB.95.035311}) and results in the renormalization of the oscillator strengths in $x$ and $y$ polarizations with 
\begin{equation}
\tilde \Gamma_{0,x} = \Gamma_0(1+P_l), \quad \tilde \Gamma_{0,y} = \Gamma_0(1-P_l),
\end{equation}
where $\Gamma_0$ is the exciton decay rate in the free space of unstrained ML~\cite{glazov2014exciton,PhysRevLett.123.067401}:
\begin{equation}
\label{Gamma0}
\Gamma_0= \frac{2\pi q}{\hbar} \left|\frac{e\hbar\gamma_3 {\Phi}(0)}{\omega_0} \right|^2
\end{equation}
with $q=\omega_0/c$ being the wave vector of exciton-induced radiation, $\omega_0$ is the exciton resonance frequency, ${\Phi}(0)$ is the electron-hole envelope function at the coinciding coordinates. In the presence of a substrate with the amplitude reflection coefficient $r_b$ (cf. Refs.~\cite{PhysRevLett.123.067401,prazdnichnykh2020control}), the splitting between $x$- and $y$-polarized excitons due to the long-range exchange interaction reads
\begin{equation}
\label{splitting:LR}
{\Delta=}\hbar \mathcal B_{lr}  (u_{xx}- u_{yy}) = \hbar\Gamma_0 P_l \Im\{r_b\}.
\end{equation}
Since $|r_b|\leqslant 1$ and $P_l \leqslant 1$ by definition, the splitting is smaller than the exciton radiative decay rate to the vacuum, $\hbar\Gamma_0$. The ratio of the splitting and the radiative linewidth in the structure with the substrate $\hbar\Gamma_0 (1+\Re\{r_b\})$~\cite{PhysRevLett.123.067401} can be evaluated as
\begin{equation}
\label{lr:ratio}
\left| \frac{\hbar \mathcal B_{lr}  (u_{xx}- u_{yy})}{\hbar\Gamma_0 (1+\Re\{r_b\})} \right| = \left|P_l\frac{\sin{\phi}}{|r_b|^{-1} + \cos{\phi}}\right|,
\end{equation}
where $\phi$ is the phase of the substrate reflection coefficient{, $r_b = |r_b|\exp{(\mathrm i \phi)}$}. Interestingly, this ratio can be sufficiently large if the substrate reflection coefficient $r_b$ is close to $-1$ ($\phi \approx \pi$). In this case the radiative linewidth of the exciton is strongly suppressed making the ratio of the splitting to the radiative linewidth in Eq.~\eqref{lr:ratio} arbitrarly large. 

For completeness, we present here the expression for the $K$-dependent long-range exchange contribution for a ML in a vacuum~\cite{glazov2014exciton,prazdnichnykh2020control}. The parameter $\mathcal A(K)$ in the effective magnetic field~\eqref{Omega} reads
\begin{equation}
\label{alpha}
\mathcal A(K) = \frac{\Gamma_0}{q\sqrt{K^2-q^2}}.
\end{equation}
We note that for the states within the light cone, $K<q$, $\mathcal A(K)$ is imaginary and the exchange interaction results in the renormalization of the exciton radiative decay rates (this part is shown in Fig.~\ref{fig:3} by the shaded grey area), while for $K>q$ the long-range exchange interaction gives rise to the splitting of the exciton states.

\subsection{Short-range exchange interaction}\label{subsec:sr}

The second contribution to the splitting results from the short-range exchange interaction which describes the Coulomb-induced mixing of Bloch functions within the unit cell~\cite{BP_exch71,birpikus_eng}.

We write the short-range exchange interaction Hamiltonian in the form
\begin{equation}
\label{SR}
\mathcal H_{sr} = \delta(\bm \rho_e - \bm \rho_h) \mathcal E_0 a_0^2 \hat U,
\end{equation}
where $a_0$ is the lattice constant, $\mathcal E_0$ is the energy parameter related to the matrix element of the Coulomb interaction calculated on the Bloch functions, and $\hat U$ is the dimensionless matrix with non-zero elements between the excitonic Bloch functions of the same symmetry, but belonging to the different valleys. 

Following Ref.~\cite{PhysRevB.95.035311} we arrive at the following expression for the splitting
\begin{equation}
\label{splitting:SR}
\Delta=\hbar\mathcal B_{sr}  (u_{xx}- u_{yy}) = (u_{xx}- u_{yy}) \mathcal E_0 a_0^2 |\Phi(0)|^2 U.
\end{equation}
Here, as above, $\Phi(\bm \rho)$ is the envelope of the electron-hole relative motion in the exciton, and the constant $U$ can be expressed via the matrix elements of $\hat U$ multiplied by the factors like $\Xi_{c,c+2}/(E_c - E_{c+2})$, \ldots, cf. Eq. (17) of Ref.~\cite{PhysRevB.95.035311}. Crude estimate for $U$ is $U \sim 1\ldots 10$. For $\mathcal E_0 = 1$~eV, $a_0=3$~\AA, and exciton Bohr radius $a_B=15$~\AA~[within the hydrogenic model $|\Phi(0)|^2=2/(\pi a_B^2)$], we obtain $\hbar\mathcal B_{sr}$ in the range of $10\ldots 100$~meV.

This mechanism of the mixing is quite analogous to that considered in Refs.~\cite{gourdon92,pikus94} for conventional quantum wells. In that case, elastic strain mixes the states of heavy and light holes (similarly to Eqs.~\eqref{Bloch:strain}), while the short-range exchange interaction provides the splitting of exciton levels. On the one hand, for quantum wells the splitting between the heavy and light holes is considerably smaller ($10\ldots 100$~meV) as compared to the interband splitting in TMDC MLs. On the other hand, the exchange interaction is much larger in TMDC MLs as compared to the quantum wells. That is why the overall effect of strain on the exciton levels is expected to be comparable in the two systems.

\subsection{Effects of non-centrosymmetricity}\label{subsec:ncs}

The parameters $\mathcal A$ and $\mathcal B$ in the effective magnetic field $\bm \Omega$, Eq.~\eqref{Omega}, are non-zero in the centrosymmetric model. Correspondingly, the mechanisms described above do not require the lack of the inversion center. Indeed, these parameters are non-zero in the centrosymmetric model where $\gamma_5$ and $\gamma_6$ in the effective Hamiltonian~\eqref{H4+} vanish, and $\Xi_{c,v}$, $\Xi_{c+2,v}$ and $\Xi_{v-3,v}$ in the strain-induced Hamiltonian~\eqref{H4+:s} vanish as well: In that case the $\bm k$-linear coupling between the Bloch states whose angular momentum components differ by two, e.g., between $c$ and $c+2$, $v-3$ is absent and the strain-induced coupling of the Bloch states with angular momentum components differ by one, e.g., between $v$ and all other bands, is absent as well. Additional contributions to the strain-induced splitting characterized by the parameter $\mathcal B$ may arise, beyond centrosymmetric approximation, e.g., due to the piezoelectric effect~\cite{Alyoruk:2015vi,Peng_2018}. Indeed, an elastic strain gives rise to a dielectric polarization~\cite{PhysRevB.104.085405} and, consequently, an in-plane electric field $\bm E = (E_x, E_y)$, with $E_x \propto 2u_{xy}$, $E_y \propto u_{xx} - u_{yy}$ that modifies an intrinsic mixing of $s$- and $p$-shell excitons studied in Ref.~\cite{PhysRevB.95.035311}. We expect, however, that such mechanisms provide weaker contributions as compared to those discussed above. Indeed, for such mechanisms the lack of the inversion center should be taken into account twice: firstly, in the piezo-effect and, secondly, in the intrinsic $s-p$-exciton mixing.

By contrast, the parameter $\mathcal C$ in Eq.~\eqref{Omega} that describes $\bm K$-linear pseudomagnetic field appears only in the non-centrosymmetric model. Evaluating the matrix elements of the four-band Hamiltonian~\eqref{H4+} on the Bloch functions~\eqref{Bloch:strain} we obtain the linear in the wavevector and strain contributions to the conduction and valence band dispersion in the $\bm K_+$ valley:
\begin{subequations}
\label{k:linear+}
\begin{multline}
E_+^c(\bm k) = \langle \tilde c|\mathcal H_+|\tilde c\rangle = E_c+ \Xi_c (u_{xx}+ u_{yy}) \\
+ [(u_{xx} - u_{yy})k_x +2u_{xy}k_y]\gamma_3\zeta_c,
\end{multline}
\begin{multline}
E_+^v(\bm k) = \langle \tilde v|\mathcal H_+|\tilde v\rangle = E_v+ \Xi_v (u_{xx}+ u_{yy}) \\
+ [(u_{xx} - u_{yy})k_x +2u_{xy}k_y]\gamma_3\zeta_v,
\end{multline}
\end{subequations}
where
\begin{subequations}
\label{zeta:cv}
\begin{align}
\zeta_c = \frac{2}{\gamma_3}\left(\frac{\gamma_6^* \Xi_{c+2,c}}{E_c - E_{c+2}} + \frac{2 \gamma_5 \Xi_{v-3,c}}{E_c - E_{v-3}}\right),\\
\zeta_v = \frac{2}{\gamma_3}\left(\frac{ \gamma_4^* \Xi_{c+2,v}}{E_v - E_{c+2}} + \frac{2 \gamma_2 \Xi_{v-3,v}}{E_c - E_{v-3}}\right).
\end{align}
\end{subequations}
The factor $\gamma_3$ is introduced in Eqs.~\eqref{k:linear+} and \eqref{zeta:cv} for convenience to make $\zeta_{c,v}$ dimensionless. We note that similarly to the parameter $\eta$ in Eq.~\eqref{eta}, the parameters $\zeta_{c,v}$ are real for consistent choice of phases of Bloch functions. In the $\bm K_-$ valley the dispersions acquire similar form, but, owing to the time-reversal, the $\bm k$-linear terms have opposite signs:
\begin{subequations}
\label{k:linear-}
\begin{multline}
E_-^c(\bm k) = E_c+ \Xi_c (u_{xx}+ u_{yy}) \\
- [(u_{xx} - u_{yy})k_x +2u_{xy}k_y]\gamma_3\zeta_c,
\end{multline}
\begin{multline}
E_-^v(\bm k) =  E_v+ \Xi_v (u_{xx}+ u_{yy}) \\
- [(u_{xx} - u_{yy})k_x +2u_{xy}k_y]\gamma_3\zeta_v.
\end{multline}
\end{subequations}
Equations~\eqref{k:linear+} and \eqref{k:linear-} describe $\bm k$-linear terms arising on the electron and hole dispersion in the presence of strain. Similarly to classical semiconductors such terms are possible only in non-centrosymmetric media~\cite{pikusmarushaktitkov_eng}. Indeed, the parameters $\zeta_c$, $\zeta_v$ contain $\gamma_5$, $\gamma_6$ and $\Xi_{c+2,v}$, $\Xi_{v-3,v}$ that vanish in the centrosymemtric model.

We can now combine Eqs.~\eqref{k:linear+} and \eqref{k:linear-} to obtain the $\Omega_z$ in Eq.~\eqref{Omega}. Taking into account that the electron, $\bm k_e$, and hole, $\bm k_h$, wavevectors are related to the exciton center of mass, $\bm K$, and relative motion, $\bm \kappa$, wavevectors by $\bm k_e = (m_e/M)\bm K - \bm \kappa$ and $\bm k_h = (m_h/M)\bm K+ \bm \kappa$ with $m_{e,h}$ being the electron and hole effective masses ($M=m_e+m_h$), and making use of the fact that the hole state is obtained from the unoccupied state in the valence band by the time reversal, we obtain
\begin{equation}
\label{C:mirco}
\mathcal C = 2\frac{\gamma_3}{\hbar} \left(\frac{m_e}{M}\zeta_c + \frac{m_h}{M}\zeta_v\right).
\end{equation}
For crude estimates the factor in parentheses can be taken as unity making $\mathcal C \sim \gamma_3/\hbar$.

\section{Experiment}\label{sec:exper}

\begin{figure}[]
	\includegraphics[width=0.85\linewidth]{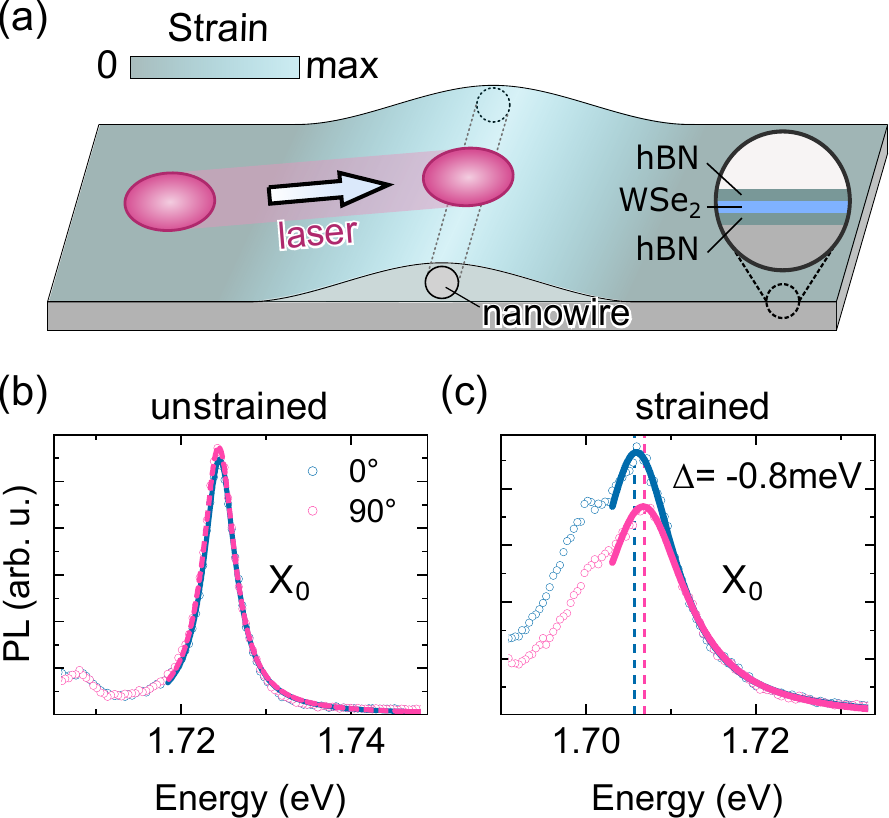}
	\caption{
	(a)~Schematic illustration of the studied hBN-encapsulated $\textrm{WSe}_2$ monolayer mechanically deformed by a single nanowire. 
	PL spectra from $X_0$ excitons are recorded at $T$=4\,K across (b) unstrained and (c) uniaxially strained regions. 
	In the strained region, the PL is shifted to lower energies and linearly polarized in parallel ($0^\circ$) and perpendicular ($90^\circ$) directions with respect to the nanowire.
	It exhibits a splitting $\Delta$ of the peak energy between the two polarizations.
	}\label{fig:4}
\end{figure}

In this section, we illustrate the impact of the strain-modified exciton dispersion shown in Fig.~\ref{fig:3} and induced linear polarization on the exciton emission in micro-photoluminescence ($\mu$-PL) experiments. 
The elastic deformations and local strains are created by transferring hBN-encapsulated $\textrm{WSe}_2$ monolayers onto $\textrm{SiO}_2/\textrm{Si}$ substrates supporting semiconductor nanowires.
A detailed description of the sample fabrication process and the characterization of the resulting uniaxial tensile strain is presented in our previous study~\cite{Florian:ub}.
These heterostructures are studied by monitoring the emission of bright $X_0$ excitons in position- and polarization-resolved $\mu$-PL scans, as schematically illustrated in Fig.~\ref{fig:4}(a). 
All measurements are conducted at $T$=4\,K to take advantage of narrow spectral linewidths.
We use a continuous-wave laser with photon energy of 2.33\,eV to excite WSe$_2$ samples non-resonantly.
These conditions are chosen to avoid optically induced valley coherence for clean analysis of linear polarization of the PL.
The laser beam is focused onto a spot diameter of $\sim1\mu$m by a 100$\times$ microscope objective. The PL spot being on the same spatial scale as the deformation by the nanowire creates additional broadening due to the averaging areas with different strain.
A combination of a half-wave plate and a linear polarizer placed in front of the detector is used to analyze the PL polarization.

PL spectra of a WSe$_2$ monolayer obtained from flat, unstrained regions of the sample show an emission peak from neutral, bright A-excitons ($X_0$) at 1.725\,eV.
Since bright $X_0$ excitons in $\bm K_\pm$ valleys emit circularly polarized light in the absence of strain [cf. Fig.~\ref{fig:1}(a)], their PL does not change when probed along two orthogonal polarization axes [see Fig.~\ref{fig:4}(b)]. 
The same measurement shows a very different result when we detect the $X_0$ emission in a strained region on top of the nanowire. 
The PL maximum shifts by $\sim$20\,meV to lower energies due to uniaxial tensile strain.
Moreover, the $X_0$ peak splits into a doublet for linear polarization parallel ($0^\circ$) and perpendicular ($90^\circ$) to the nanowire axis, i.e., along the expected main axis of the strain tensor, in accordance with the theoretical predictions, see Figs.~\ref{fig:1} and \ref{fig:3}. The low energy shoulder visible in both polarizations shifts accordingly for the two polarizations and is attributed to a region with slightly higher strain.
The maximum splitting is on the order of 1\,meV and the low-energy peak emits at a higher intensity corresponding to a polarization degree of about 15\,\%.

\begin{figure}[]
	\includegraphics[width=0.85\linewidth]{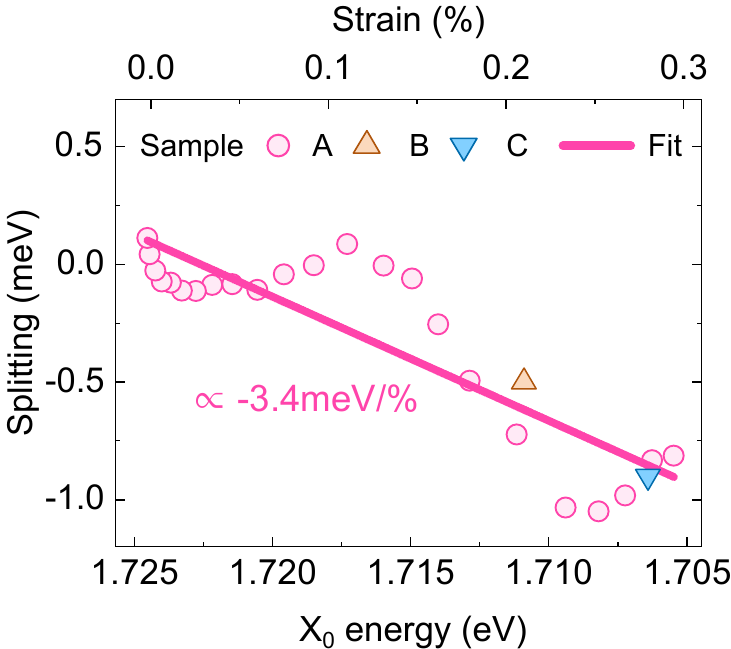}
	\caption{Linear-polarization energy splitting $\Delta$ extracted from a polarization-resolved $\mu$-PL line scan across the nanowire-induced deformation is shown as a function of the $X_0$ energy (bottom axis) and the equivalent values of strain (top axis). Uniaxial tensile strain is calibrated by $-63.2$~meV per $\%$ strain. A line fit estimates the strain dependence of the energy splitting $\Delta$ to be $-3.4$~meV per $\%$ strain. Experimental observations in sample A are confirmed by polarization-resolved measurements on two other nanowire-induced deformations denoted sample B (orange triangle) and C (blue triangle).}\label{fig:5}
\end{figure}

Figure~\ref{fig:5} presents the energy splitting $\Delta$ obtained by scanning the laser excitation spot across the elastically deformed region and fitting the recorded $X_0$ emission for both polarization axes by Gaussian functions.
While the absolute values of the energy splittings are comparatively small, we observe a consistent dependence of the splitting on the position of the sample and on the energy of the PL emission maximum. 
We note that the additional oscillatory behavior superimposed on $\Delta$ is likely related to a residual coupling between $X_0$ excitons and the optical modes of the nanowire waveguide-like structure, see below. Other common sources of the exciton energy shift, such as spatial fluctuations of the doping density or of the dielectric environment, are unlikely to contribute since neither the trion intensity nor the binding energy of the ground state exciton changes significantly, as shown in our previous study~\cite{Florian:ub}. Measurements of $\Delta$ in two other samples quantitatively confirm the results presented in Figs.~\ref{fig:4} and \ref{fig:5}. 
A linear fit to the data provides a scaling factor of $-3.4$\,meV per $\%$ uniaxial tensile strain compared to $-63.2$\,meV per $\%$ strain for the absolute shift of the exciton peak with the latter value being consistent with the exciton deformation potentials in WSe$_2$ MLs~\cite{2053-1583-3-2-021011,Florian:ub}.

The observed strain induced linear-polarization splitting of the exciton is in accord with the theory developed in Sec.~\ref{sec:micro}. As we have shown, there are two contributions to the strain-induced splitting: due to the long-range exchange interaction, Eq.~\eqref{splitting:LR}, and due to the short-range exchange interaction, Eq.~\eqref{splitting:SR}. According to the estimate using Eq.~\eqref{splitting:SR} one can expect the short-range contribution to $\Delta \sim 1$~meV per $\%$ strain. Note that for $\hbar\Gamma_0$ in the range of several meV and $P_l \lesssim 15\%$ the long-range exchange interaction provides by far smaller contribution to the splitting $\Delta$ as compared to the experimentally observed one in Fig.~\ref{fig:5}. It can, however, at least in principle, explain oscillations of the splitting superimposed over the linear behavior: Indeed, in our sample the interference of light reflected from the substrate and the nanowire can result in position-dependent oscillations of the substrate reflectivity $r_b$ [cf. Ref.~\cite{PhysRevLett.123.067401}] which yield oscillations of $\Delta$ in Eq.~\eqref{splitting:LR} as a function of coordinate and, consequently, of the strain, see Fig.~\ref{fig:4}(a). The values of the induced linear polarization are also consistent with theoretical prediction, Eq.~\eqref{Pl}~\footnote{For tensile strain $u_{xx} -u_{yy}>0$ with $y$-axis being the nanowire axis in Fig.~\ref{fig:4}(a). In experiment, $P_l<0$, light is predominantly polarized along the nanowire, see Fig.~\ref{fig:4}(b), thus we can infer that the parameter $\eta<0$ in Eq.~\eqref{Pl}. Similarly, from the sign of $\Delta<0$, Fig.~\ref{fig:5}, we obtain $\mathcal B_{sr}<0$.}. We abstain from a more detailed quantitative comparison between the model and experiment which requires (i) precise atomistic calculations of the interband deformation potentials and short-range exchange interaction parameters, (ii) experimental separation of the waveguide effects on the strain-induced splitting $\Delta$, and (iii) analysis of temperature effect on the linear polarization degree to separate overall polarization due to the variation of the optical selection rules and induced polarization due to the exciton thermalization over the strain-split fine structure states.

The observed behavior of the exciton resonance in mechanically-deformed WSe$_2$ samples thus provides an experimental illustration for the fine structure splitting of the radiative doublet under strain.

\section{Conclusion and outlook}\label{sec:concl}

We have developed a theory of strain-induced effects on the exciton energy spectrum fine structure in 2D TMDCs. We have shown that the elastic strain produces optical anisotropy of a monolayer resulting in deviation from the chiral selection rules. The induced polarization is linked to the main axes of the strain tensor. Further, the strain gives rise to a wavevector independent polarization splitting of the exciton radiative doublet, experimentally demonstrated for uniaxially strained WSe$_2$ monolayers. We have analyzed an interplay of this splitting with the wavevector dependent longitudinal-transverse splitting of excitonic states and shown that these contributions can compensate each other for particular values of the exciton wavevector. In the vicinity of these compensation points the exciton energy spectrum has a conical, massless Dirac, form. We have also identified a strain-induced wavevector dependent contribution to the exciton splitting in circular polarizations linked with the specific three-fold rotational symmetry of TMDC monolayers with broken spatial inversion. 
Microscopic mechanisms behind the strain-related contributions to the exciton effective Hamiltonian have been identified. The microscopic model has been developed within the four-band $\bm k\cdot \bm p$-Hamiltonian, and the estimates of all relevant contributions are presented.

The developed theory provides a basis for the strain tuning of exciton fine structure~\cite{Chakraborty2020} (see also Refs.~\cite{PhysRevB.88.155330,Su:2016ul} for conventional semiconductor systems), an important property for quantum technology and the application of exciton and biexciton emission for generation of entagled photons~\cite{shields06}. The formation of a Dirac-like dispersion may be useful for the relalization of topological-like effects in excitonic systems~\cite{PhysRevLett.114.116401,Klembt:2018aa,PhysRevApplied.17.024037}.
 Strain-induced spatially inhomogeneous pseudomagnetic fields may also be realized in bilayer structures where moir\'e effects are important~\cite{Seyler:2019aa,Tran:2019aa,Jin:2019aa,Alexeev:2019aa} enabling adiabatic exciton pseudospin evolution. Strain-induced exciton fine structure splitting can be also important for the optomechanical applications of excitons in atomically-thin crystals~\cite{Morell:2019aa,Avdeev:2020aa,Xie:2021wv,2022arXiv220212143I}.

\acknowledgements
Financial support of the theoretical work by M.M.G. via RSF project 19-12-00051 is gratefully acknowledged.
Work at CUNY was supported through the NSF QII TAQS 1936276 and NSF DMR-2130544 grants.
We further acknowledge financial support by the DFG via Emmy Noether Initiative (CH 1672/1, Project-ID: 287022282), as well as Project ID 422 31469
5032-SFB1277 (subprojects A01 and B05), and the W\"urzburg-Dresden Cluster of Excellence on Complexity and Topology in Quantum Matter ct.qmat (EXC 2147, Project-ID 390858490).
K.W. and T.T. acknowledge support from JSPS KAKENHI (Grant Numbers 19H05790, 20H00354 and 21H05233).

\end{document}